\documentstyle[twoside,epsfig]{article}

\catcode`\@=11
\long\def\@makefntext#1{
\protect\noindent \hbox to 3.2pt {\hskip-.9pt  
$^{{\eightrm\@thefnmark}}$\hfil}#1\hfill}		

\def\thefootnote{\fnsymbol{footnote}}
\def\@makefnmark{\hbox to 0pt{$^{\@thefnmark}$\hss}}	
	
\def\ps@myheadings{\let\@mkboth\@gobbletwo
\def\@oddhead{\hbox{}
\rightmark\hfil\eightrm\thepage}   
\def\@oddfoot{}\def\@evenhead{\eightrm\thepage\hfil
\leftmark\hbox{}}\def\@evenfoot{}
\def\sectionmark##1{}\def\subsectionmark##1{}}

\oddsidemargin=\evensidemargin
\renewcommand{\thefootnote}{\fnsymbol{footnote}}

\newcounter{sectionc}\newcounter{subsectionc}\newcounter{subsubsectionc}
\renewcommand{\section}[1] {\vspace{12pt}\addtocounter{sectionc}{1} 
\setcounter{subsectionc}{0}\setcounter{subsubsectionc}{0}\noindent 
	{\tenbf\thesectionc. #1}\par\vspace{5pt}}
\renewcommand{\subsection}[1] {\vspace{12pt}\addtocounter{subsectionc}{1} 
	\setcounter{subsubsectionc}{0}\noindent 
	{\bf\thesectionc.\thesubsectionc. {\kern1pt \bfit #1}}\par\vspace{5pt}}
\renewcommand{\subsubsection}[1] {\vspace{12pt}\addtocounter{subsubsectionc}{1}
	\noindent{\tenrm\thesectionc.\thesubsectionc.\thesubsubsectionc.
	{\kern1pt \tenit #1}}\par\vspace{5pt}}
\newcommand{\nonumsection}[1] {\vspace{12pt}\noindent{\tenbf #1}
	\par\vspace{5pt}}

\newcounter{appendixc}
\newcounter{subappendixc}[appendixc]
\newcounter{subsubappendixc}[subappendixc]
\renewcommand{\thesubappendixc}{\Alph{appendixc}.\arabic{subappendixc}}
\renewcommand{\thesubsubappendixc}
	{\Alph{appendixc}.\arabic{subappendixc}.\arabic{subsubappendixc}}

\renewcommand{\appendix}[1] {\vspace{12pt}
        \refstepcounter{appendixc}
        \setcounter{figure}{0}
        \setcounter{table}{0}
        \setcounter{lemma}{0}
        \setcounter{theorem}{0}
        \setcounter{corollary}{0}
        \setcounter{definition}{0}
        \setcounter{equation}{0}
        \renewcommand{\thefigure}{\Alph{appendixc}.\arabic{figure}}
        \renewcommand{\thetable}{\Alph{appendixc}.\arabic{table}}
        \renewcommand{\theappendixc}{\Alph{appendixc}}
        \renewcommand{\thelemma}{\Alph{appendixc}.\arabic{lemma}}
        \renewcommand{\thetheorem}{\Alph{appendixc}.\arabic{theorem}}
        \renewcommand{\thedefinition}{\Alph{appendixc}.\arabic{definition}}
        \renewcommand{\thecorollary}{\Alph{appendixc}.\arabic{corollary}}
        \renewcommand{\theequation}{\Alph{appendixc}.\arabic{equation}}
        \noindent{\tenbf Appendix \theappendixc #1}\par\vspace{5pt}}
\newcommand{\subappendix}[1] {\vspace{12pt}
        \refstepcounter{subappendixc}
        \noindent{\bf Appendix \thesubappendixc. {\kern1pt \bfit #1}}
	\par\vspace{5pt}}
\newcommand{\subsubappendix}[1] {\vspace{12pt}
        \refstepcounter{subsubappendixc}
        \noindent{\rm Appendix \thesubsubappendixc. {\kern1pt \tenit #1}}
	\par\vspace{5pt}}

\newcommand{\textlineskip}{\baselineskip=13pt}
\newcommand{\smalllineskip}{\baselineskip=10pt}

\def\eightcirc{
\begin{picture}(0,0)
\put(4.4,1.8){\circle{6.5}}
\end{picture}}
\def\eightcopyright{\eightcirc\kern2.7pt\hbox{\eightrm c}}

\def\abstracts#1#2#3{{
	\centering{\begin{minipage}{4.5in}\baselineskip=10pt\footnotesize
	\parindent=0pt #1\par 
	\parindent=15pt #2\par
	\parindent=15pt #3
	\end{minipage}}\par}} 

\newcommand{\bibit}{\nineit}

\renewenvironment{thebibliography}[1]
	{\frenchspacing
	 \ninerm\baselineskip=11pt
	 \begin{list}{\arabic{enumi}.}
	{\usecounter{enumi}\setlength{\parsep}{0pt}
	 \setlength{\leftmargin 12.7pt}{\rightmargin 0pt} 
	 \setlength{\itemsep}{0pt} \settowidth
	{\labelwidth}{#1.}\sloppy}}{\end{list}}

\newcounter{itemlistc}
\newcounter{romanlistc}
\newcounter{alphlistc}
\newcounter{arabiclistc}

\newcommand{\fcaption}[1]{
        \refstepcounter{figure}
        \setbox\@tempboxa = \hbox{\footnotesize Fig.~\thefigure. #1}
        \ifdim \wd\@tempboxa > 5in
           {\begin{center}
        \parbox{5in}{\footnotesize\smalllineskip Fig.~\thefigure. #1}
            \end{center}}
        \else
             {\begin{center}
             {\footnotesize Fig.~\thefigure. #1}
              \end{center}}
        \fi}

\newcommand{\tcaption}[1]{
        \refstepcounter{table}
        \setbox\@tempboxa = \hbox{\footnotesize Table~\thetable. #1}
        \ifdim \wd\@tempboxa > 5in
           {\begin{center}
        \parbox{5in}{\footnotesize\smalllineskip Table~\thetable. #1}
            \end{center}}
        \else
             {\begin{center}
             {\footnotesize Table~\thetable. #1}
              \end{center}}
        \fi}

\def\@citex[#1]#2{\if@filesw\immediate\write\@auxout
	{\string\citation{#2}}\fi
\def\@citea{}\@cite{\@for\@citeb:=#2\do
	{\@citea\def\@citea{,}\@ifundefined
	{b@\@citeb}{{\bf ?}\@warning
	{Citation `\@citeb' on page \thepage \space undefined}}
	{\csname b@\@citeb\endcsname}}}{#1}}

\newif\if@cghi
\def\cite{\@cghitrue\@ifnextchar [{\@tempswatrue
	\@citex}{\@tempswafalse\@citex[]}}
\def\citelow{\@cghifalse\@ifnextchar [{\@tempswatrue
	\@citex}{\@tempswafalse\@citex[]}}
\def\@cite#1#2{{$\null^{#1}$\if@tempswa\typeout
	{IJCGA warning: optional citation argument 
	ignored: `#2'} \fi}}

\def\pmb#1{\setbox0=\hbox{#1}
	\kern-.025em\copy0\kern-\wd0
	\kern.05em\copy0\kern-\wd0
	\kern-.025em\raise.0433em\box0}

\def\fnt#1#2{\footnotetext{\kern-.3em
	{$^{\mbox{\scriptsize #1}}$}{#2}}}

\def\fpage#1{\begingroup
\voffset=.3in
\thispagestyle{empty}\begin{table}[b]\centerline{\footnotesize #1}
	\end{table}\endgroup}

\font\tenrm=cmr10
\font\tenit=cmti10 
\font\tenbf=cmbx10
\font\bfit=cmbxti10 at 10pt
\font\ninerm=cmr9
\font\nineit=cmti9

\font\eightrm=cmr8

\def\qed{\hbox{${\vcenter{\vbox{			
   \hrule height 0.4pt\hbox{\vrule width 0.4pt height 6pt
   \kern5pt\vrule width 0.4pt}\hrule height 0.4pt}}}$}}

\renewcommand{\thefootnote}{\fnsymbol{footnote}}	

\begin{document}


\normalsize\textlineskip
\thispagestyle{empty}
\setcounter{page}{1}


\vspace*{0.88truein}
\fpage{1}
\centerline{\bf NUMERICAL ASPECTS OF BUBBLE NUCLEATION}
\vspace*{0.37truein}
\centerline{\footnotesize SURUJHDEO SEUNARINE and DOUGLAS W. MCKAY}
\vspace*{0.015truein}
\centerline{\footnotesize\it Department of Physics and Astronomy, The University
of Kansas}
\baselineskip=10pt
\centerline{\footnotesize\it Lawrence, KS 66044, USA}
\vspace*{10pt}

\vspace*{0.21truein}
\abstracts{Bubble nucleation has been studied 
on lattices using phenomenological Langevin equations. Recently there have been theoretical 
motivations for using these equations.  These studies also conclude that the simple 
Langevin description requires some modification. We study bubble nucleation  
on a lattice and determine effects of the modified Langevin equations.}{}{}


\vspace*{1pt}\textlineskip	
\section{Introduction}		
\vspace*{-0.5pt}
\noindent
 The effective potential determines the vacuum state of a quantum field theory.$^1$ 
When radiative corrections are included or, if one constructs the
finite temperature version of field theory the (effective)potential 
may have two minimum, one of which
is lower than the other. The 'higher' minimum is referred to as the false or metastable vacuum. 
The global minimum is the true vacuum state of the theory. The existence of these metastable 
vacua have interesting consequences in the early Universe.  If at early times the field settles 
in the false vacuum it may 'decay' to the true vacuum either by quantum tunneling or thermal 
hopping. The transition region is spherically symmetric in 
coordinate space.  Bubbles of the stable phase appear in the metastable phase.  
In relativistic field theory a formalsim for calculating decay rates from the unstable to the
stable vacuum has been developed.$^2$  
Since potentials which are non-linear in the fields are inherent in these problems, it is
usually not always possible to find analytic solutions to the decay rate equations. Numerical methods 
are employed in order to avoid too many analytic approximations.  
For example, in one detailed numerical study, a scalar field in 1 + 1 dimensions in
contact with a thermal bath 
was modeled using a phenomenological Langevin equation with a field independent, 
white noise driving term.$^3$ 
 There have been attempts to derive a Langevin equation containing fluctuation and dissipation
terms from purely field theory considerations.$^{4}$  In these 
works the authors considered different models for the thermal bath, but the common result 
obtained was the possiblity that the thermal noise in the Langevin equation 
could be colored and depend on the field.
In this work we investigate the effects of these non-linear fluctuation and 
dissipation terms numerically. 

\textheight=7.8truein
\setcounter{footnote}{0}
\renewcommand{\thefootnote}{\alph{footnote}}
\vspace*{1pt}\textlineskip	
\section{Numerical Method}		
\begin{figure}[hbtp]
\vspace{9pt}
\centerline{\hbox{ \hspace{0.0in} 
    \epsfxsize=2.5in
    \epsffile{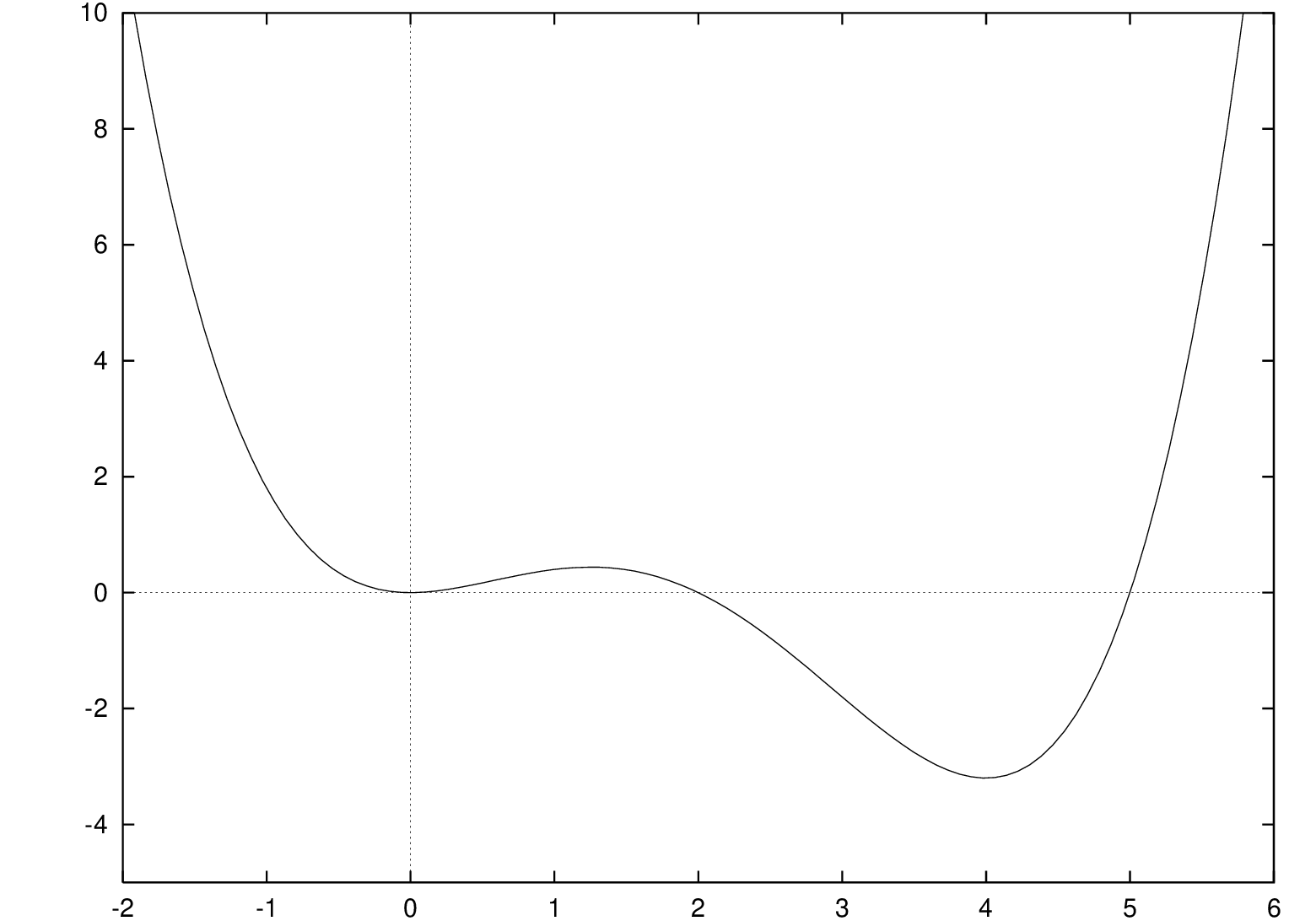}
    \hspace{0.1in}
    \epsfxsize=2.5in
    \epsffile{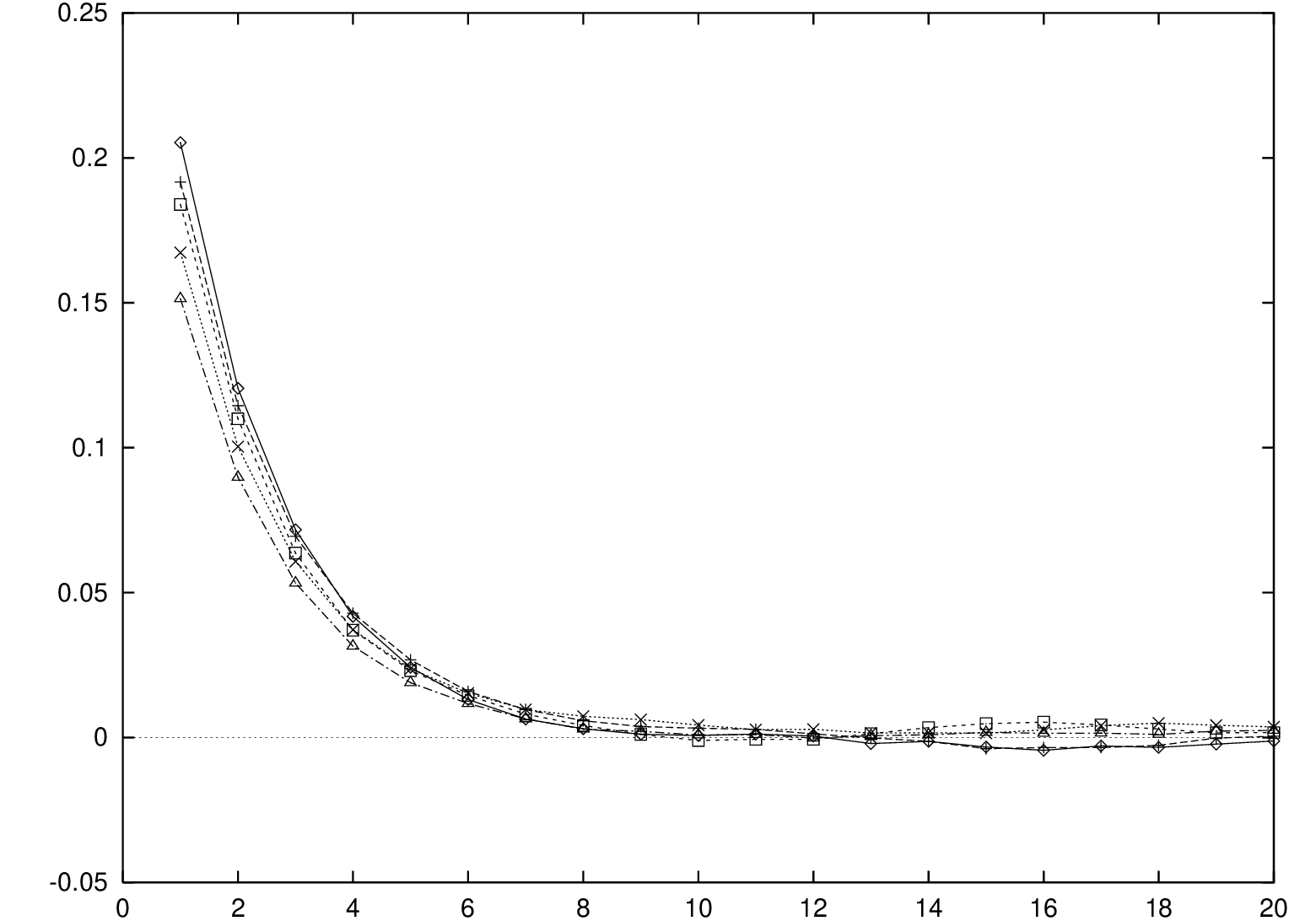}
    }
  }
\vspace{9pt}
\hbox{\hspace{1.35in} (a) \hspace{2.10in} (b)} 
\fcaption{(a)The potential $V(\phi)=\phi^2-\alpha\phi^3+\lambda\phi^4$ vs $\phi$: 
(b)The correlation $<\phi(x)\phi(x')>$ vs
$|x-x'|$ for initial conditions at different temperatures}
\label{pot_corr}
\end{figure}

\vspace*{-0.5pt}
\noindent
We study nucleation by evolving the equations of motion of the field on a lattice using
a staggerd leap-frog algorithm. We
work in $1+1$ dimensions and assume that the decay is thermally driven.  The field 
is coupled to the thermal bath through a stochastic noise term, $\xi(x,t)$.
\begin{equation}
\frac{\partial^2\phi(x,t)}{\partial x^2}-\frac{\partial^2\phi(x,t)}{\partial t^2}-
	\eta F(\phi(x,t))\frac{\partial\phi(x,t)}{\partial t}=-V(\phi)+\xi(x,t)G(\phi(x,t))
	\label{full_noise}
\end{equation}
where $F(\phi(x,t))=G(\phi(x,t))=1$ for {\it additive noise} and, $F(\phi(x,t))=\phi(x,t)^2$ and
$G(\phi(x,t))=\phi(x,t)$ for {\it multiplicative noise}.  The general shape
of the potential is show in Fig. \ref{pot_corr}(a).
The noise is assumed to be Gaussian, in terms of the distribution
of sizes of the fluctuations, and white, in that the fluctuations are uncorrelated in 
space and time. We assume the noise and viscosity terms are related through the 
fluctuation-dissipation theorem.$^{5}$
\begin{equation}
	<\xi(x,t)\xi(x',t')>=2T\eta\delta(x-x')\delta(t-t').
	\label{f_d1}
\end{equation}
In later work a generalized fluctuation-dissipation 
theorem for the multiplicative noise will be discussed.
We start with the field in the false vacuum, $V(\phi)=0$, and evolve the equation of 
motion on the lattice until the field acquires 
its true vacuum value at enough neighboring space points 
to avoid counting as bubbles fluctuations that 
re-collapse to the false vacuum value.

\vspace*{1pt}\textlineskip	
\section{Results}		
\begin{figure}[hbtp]
  \vspace{9pt}

  \centerline{\hbox{ \hspace{0.0in} 
    \epsfxsize=2.2in
    \epsffile{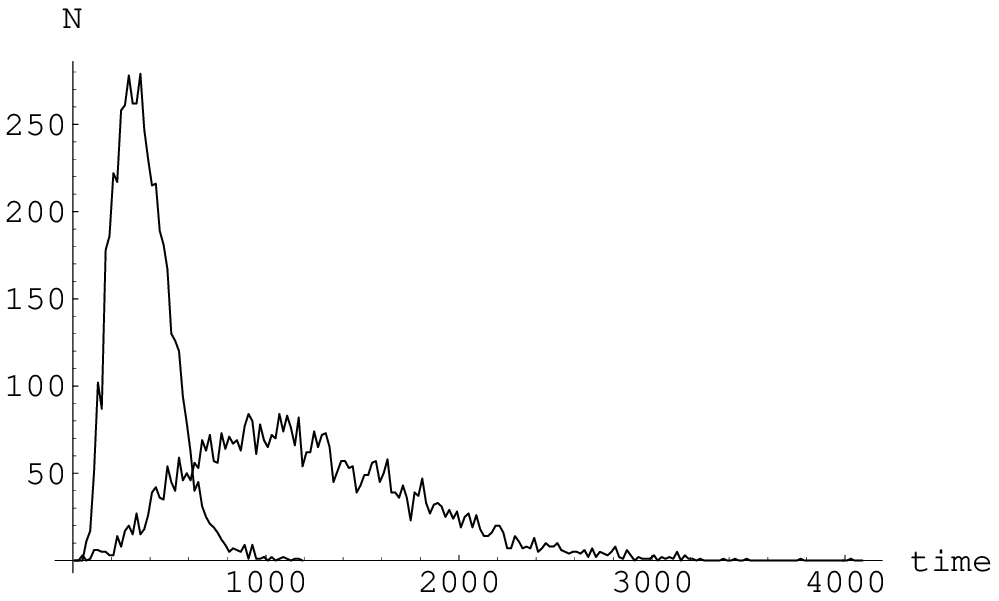}
    \hspace{0.2in}
    \epsfxsize=2.20in
    \epsffile{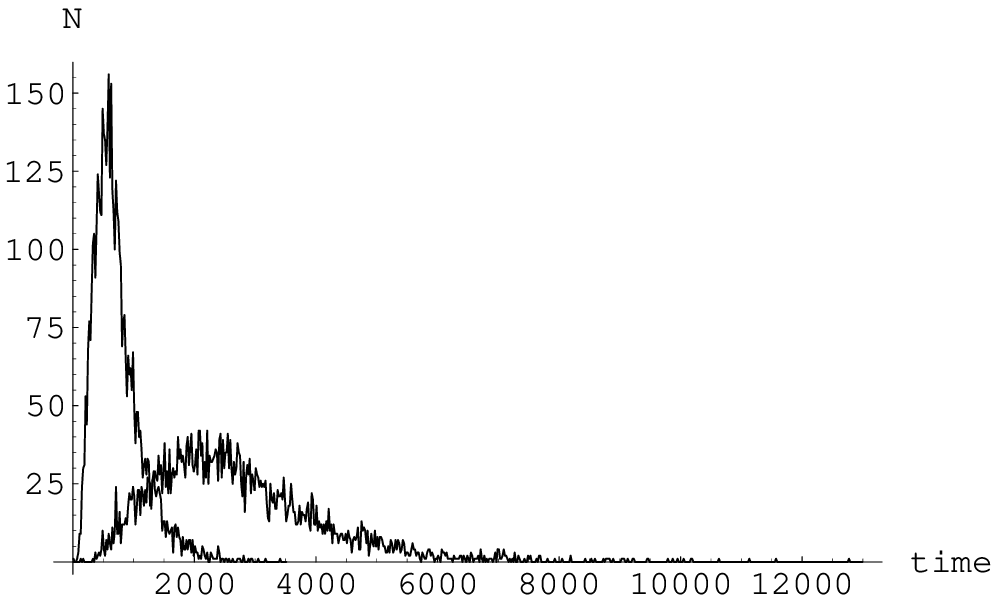}
    }
  }
\vspace{9pt}
\hbox{\hspace{1.35in} (a) \hspace{2.10in} (b)} 
\fcaption{Nucleation times at (a) T=1.6 and (b) T=1.1. Ther sharper peaks are the additive
noise cases.}
\label{times}
\end{figure}

\vspace*{-0.5pt}
\noindent
We report on some observations and preliminary results of our study. A large number 
of nucleation times has to be obtained in order to make a good estimate of the 
nucleation rate. Furthermore, the nucleation times obtained from the simulation have
to be suitably interpreted. We found that if one starts in the false vacuum with a 
uniform field configuration or with one which is random and uncorrelated then there
is a long waiting time before the any bubble is nucleated. This delay is interpreted
as the time taken for the coupled field to acquire the short range correlation that
is needed for a bubble to nucleate. Nucleation of bubbles occurs only when the field
at some point in space reaches the true vacuum and {\it pulls} along neighbouring
points.  A quenching technique, where the equations are evolved in an unbroken potential
until the short distance correlation, Fig.\ref{pot_corr}(b), is obtained, can be used to 
eliminate the delay time.  Fig. \ref{times}(a) and (b) show the nucleation times recorded
for $5000$ bubbles using separately the additive and multiplicative noise 
from (\ref{full_noise}). We used $\alpha=0.74$, $\lambda=0.1$, and $\eta=1$.  
At the time 
of quenching we reset the time to zero.  We see that the distributions are not symmetric. This is
expected for the distribution of waiting times for a decay process. We also see that the for the
multiplicative noise case the nucleation times are larger and more broadly distributed.  
Although the size of the thermal fluctuations is multiplied by the field value, the nonlinear
dissipation term has an overall larger effect and retards the decay in comparison to the 
additive noise case.  If one obtains non-linear Langevin type equations when studying realistic
models like the Electroweak effective potential, for example, the implication for the 
time scale of nucleation will require a similar careful study. Other aspects of these
non-linear models currently under study include the effects on bubble wall expansion and
the spatial distribution of critical fluctuations.    

\nonumsection{Acknowledgements}
\noindent
We thank Ruslan Davidchack, Brian Laird, John Ralston, and Hume Feldman
for very helpful discussions.


\nonumsection{References}


\begin{thebibliography}{000}
\bibitem{1}
G. Jona-Lasinio, {\bibit Nuovo Cimento}, {\bf 34}, 1790 (1964);
K. Symanzik, {\bibit Comm. Math. Phys}, {\bf 16}, 48 (1970) 

\bibitem{2}
S. Coleman, {\bibit Phys. Rev.}, {\bf D15}, 2929, (1977).

\bibitem{3}
M. Alford, H. Feldman, M. Gleiser, {\bibit Phys. Rev.}, {\bf D47}, 2168, (1993).

\bibitem{4}
M.Gleiser, R.Ramos, {\bibit Phys.Rev.}, {\bf D50}, 2441 (1994); D. Lee, D.  Boyanovsky, 
{\bibit Nucl.Phys.}, {\bf B406}, 631, (1993) 

\bibitem{5}
S. Chandrasekhar {\bibit Rev.Mod.Phys.}, {\bf 15}, 1, 1943 

\end{thebibliography}
\end{document}